\documentclass[12pt,preprint]{emulateapj}
\usepackage{epsfig}
\shorttitle{Fluorine in PNe}
\shortauthors{Zhang \& Liu}

\begin{document}

\title{Fluorine abundances in planetary nebulae}

\author{Y. Zhang\altaffilmark{1,2} and X.-W. Liu\altaffilmark{1}} 

\altaffiltext{1}{Department of Astronomy, Peking University, Beijing 100871,
        P. R. China; E-mail: zhangy@bac.pku.edu.cn.}

\altaffiltext{2}{Space Telescope Science Institute, 3700 San Martin Drive, Baltimore, MD
21218, USA.}

\begin{abstract} 
We have determined fluorine abundances from the [\ion{F}{2}] $\lambda$4789 and
[\ion{F}{4}] $\lambda$4060 nebular emission lines for a sample of planetary
nebulae (PNe). Our results show that fluorine is generally overabundant in PNe,
thus providing new evidence for the synthesis of fluorine in asymptotic giant
branch (AGB) stars. [F/O] is found to be positively correlated with the C/O
abundance ratio, in agreement with the predictions of theoretical models of
fluorine production in thermally pulsing AGB stars.  A large
enhancement of fluorine is observed in the Wolf-Rayet PN NGC\,40, suggesting
that high mass-loss rates probably favor the survival of fluorine.
\end{abstract}


\section{Introduction}

The astrophysical origin of fluorine is still an unresolved issue.  The element
has only one stable, yet rather fragile, isotope $^{19}$F.  In stellar
interiors it is readily annihilated by the most abundant elements hydrogen and
helium, via reactions $^{19}$F(p, $\alpha$)$^{16}$O and $^{19}$F($\alpha$,
p)$^{22}$Ne, respectively. In order to explain the presence of F a mechanism
is required that enables F to escape from the hot stellar interior after it
is created.  Three scenarios have been proposed as the potential sources of F:
explosions of Type {\sc ii} supernovae (SNe), stellar winds of Wolf-Rayet (WR)
stars, and the third dredge-up of asymptotic giant branch (AGB) stars.
\citet{w88} show that inelastic scattering of neutrinos escaping from the
collapsing core of a Type {\sc ii} SN by nuclei in the shell, often referred to
as the $\nu$-process, can convert $^{20}$Ne into $^{19}$F.  In WR stars,
$^{19}$F is probably produced during the He-burning phase and then ejected into
space by strong stellar winds before it is destroyed \citep{m00}. For AGB
stars, \citet{f92} propose that $^{19}$F is synthesized and then dredged up to
the surface during the He-burning thermal pulses. The reaction chain for F
production in WR and AGB stars is $^{14}$N($\alpha$,
$\gamma$)$^{18}$F($\beta^+$)$^{18}$O(p, $\alpha$)$^{15}$N($\alpha$,
$\gamma$)$^{19}$F, where protons are liberated through $^{13}$C($\alpha$,
n)$^{16}$O followed by neutron captures $^{14}$N(n, p)$^{14}$C. Based on a
semi-analytic model, \citet{r04} suggest that both WR and AGB stars are
significant sources of F. To discriminate between the three possible scenarios of F
production, accurate measurements of F abundances in these objects are needed.

Observations of F outside the solar system have been scarce.  A pioneering work
was carried out by \citet{j92}, who determined F abundances in red giants using
infrared HF vibration-rotation transitions and found enhanced F in C-rich
stars, providing evidence of F production in AGB stars. Their results were
supported by observations of \ion{F}{5} and \ion{F}{6} absorption lines in the
far-UV spectra of hot post-AGB stars \citep{w05}.  On the other hand,
\citet{c03} measured HF lines for a sample of nine red giants (RGs) in the Large
Magellanic Cloud (LMC) and two RGs in the Galactic globular cluster $\omega$
Centauri. Very low F/O ratios were found for the two metal-poor $\omega$ Cen
giants. Given that the cluster must have been significantly enriched by AGB
evolution, as indicated by large enhancements of $s$-process elements observed
in its member stars, they concluded that AGB stars are probably not the
dominant source of F. To our knowledge, no observation of F in a WR star has
been reported to date. The first measurement of interstellar F was obtained by
\citet{f05}, who detected \ion{F}{1} $\lambda$955 interstellar absorption in
two sight lines towards the Cep OB2 association using the Far Ultraviolet
Spectroscopic Explorer ({\it FUSE}).  Their measurements yield no evidence of
enhanced F abundances resulting from the $\nu$-process in Type {\sc ii} SNe. 

Planetary nebulae (PNe) are ejected from AGB stars and they therefore
serve as a direct test
bed for F production in AGB stars. Given its low condensation temperature,
depletion of F onto dust grains is unlikely to be significant in PNe.
Nevertheless, due to its intrinsic low abundance, F emission lines are
difficult to measure in PN spectra.  Based on uncertain measurements of the
[\ion{F}{4}] $\lambda$4060 nebular line in two PNe, \citet{a83} claimed that the
F abundance in PNe was consistent with the solar value. A definite measurement of
the F abundance was presented by \citet{l98} for the high-excitation PN NGC\,4361,
indicating that F was overabundant in this metal-poor PN possibly belonging to
the halo population.

In this paper we present F abundances for a sample of PNe and we examine the F
production in AGB stars. In Section 2, we survey measurements of the
[\ion{F}{2}] $\lambda$4789 and the [\ion{F}{4}] $\lambda$4060 lines available
from recently published high quality spectra. The derived abundances are
presented in Section 3.  A discussion follows in Section 4, and section 5
summarizes the results.

\begin{figure*}
\centering
\epsfig{file=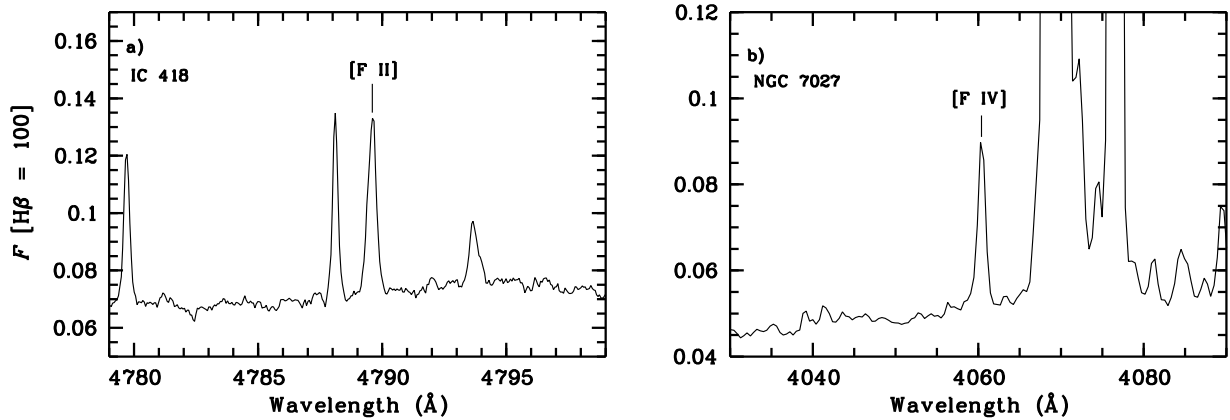, height=5.5cm}
\caption{a) Spectrum of a low-excitation PN IC\,418 showing the
[\ion{F}{2}] $\lambda$4789 line; b) Spectrum of a high-excitation
PN NGC\,7027 showing the [\ion{F}{4}] $\lambda$4060 line. Note that
interstellar extinction has not been corrected for.}
\label{spe}
\end{figure*}

\section{Fluorine emission lines}

Among the ionic species of fluorine, F$^+$ and F$^{3+}$ are detectable in the
optical via the [\ion{F}{2}] $^1$D$_2$--$^3$P$_2$ $\lambda4789.48$ and the
[\ion{F}{4}] $^1$D$_2$--$^3$P$_2$ $\lambda4059.94$ nebular lines, respectively.
We have surveyed high signal-to-noise spectra published in the literature (see
the last column of Table~\ref{F_PN}).  Many of them are published recently by
our own and other research groups, obtained in spectroscopic surveys aimed at
detecting weak emission lines from heavy element ions. 

PNe from which the [\ion{F}{2}] $\lambda4789$ line or the [\ion{F}{4}]
$\lambda4060$ line have been detected are listed in Table~\ref{F_PN}.  The fact
that all PNe where the [\ion{F}{2}] line is detectable are low-excitation PNe
and all PNe from which the [\ion{F}{4}] has been detected are high-excitation
objects strengthens the confidence in our line identifications.  Fig.~\ref{spe}
plots the spectra of two PNe, one of low-excitation and another of
high-excitation, around the wavelengths of the [\ion{F}{2}] line and the
[\ion{F}{4}] line, respectively. The excitation class (E.C.) and the dereddened
intensities of the [\ion{F}{2}] $\lambda4789$ and the [\ion{F}{4}]
$\lambda4060$ line, relative to that of H$\beta$, are listed respectively in
Cols. 2, 3, and 4 of Tables~\ref{F_PN}. Typical intensities of the two fluorine
lines relative to H$\beta$ are of the order of $10^{-4}$. 

\citet{b00} report detection of the [\ion{F}{2}] $\lambda4789$ line from the
Orion Nebula, albeit with a low signal-to-noise ratio. A feature at the right
wavelength has also been detected by \citet{p04} in the low-excitation PN
NGC\,5315, but identified by them as \ion{N}{2}. Given that the feature
does not appear in high-excitation PNe, we regard this as a mis-identification.
The detection of this line is very uncertain in the low-excitation PN IC\,2501
because of severe contamination at this wavelength by scattered light from
H$\beta$.

\begin{deluxetable*}{lcccccccl}
\tabletypesize{\scriptsize}
\tablecaption{Measurements of fluorine in nebulae\tablenotemark{a}.
\label{F_PN}}
\tablewidth{0pt}
\tablehead{
\colhead{Source} & \colhead{E.C.$^b$} &
\colhead{$I$([\ion{F}{2}]$_{4789}$)$^c$} &
\colhead{F$^+$/H$^+$} &
\colhead{O$^+$/H$^+$}&
\colhead{F/O$^d$} & \colhead{C/O}&
\colhead{O/H} &\colhead{Ref.}
}
\startdata
IC\,418  &   1   & 2.72$(-4)$ &  4.34$(-8)$ &  1.70$(-4)$ & 2.55$(-4)$&1.8   &2.90$(-4)$ & S04 \\
IC\,2501 &   3   & 1.70$(-5)$:&  3.38$(-9)$:&  8.25$(-5)$ & 4.09$(-5)$:&1.2   &5.13$(-4)$ & W05a\\
NGC\,40  &   1   & 6.60$(-4)$ &  2.95$(-7)$ &  4.83$(-4)$ & 6.10$(-4)$&1.4   &4.90$(-4)$ & L04 \\
NGC\,5315&   2   & 4.40$(-5)$ &  6.45$(-9)$ &  4.17$(-5)$ & 1.54$(-4)$&0.95  &7.41$(-4)$ & P04 \\
Orion    &\nodata& 5.37$(-5)$ &  9.01$(-9)$ &  7.94$(-5)$ & 1.13$(-4)$&0.58  &4.47$(-4)$ & B00 \\
\\[-0.06in]
\hline
\hline
\\[-0.07in]
\colhead{Source} & \colhead{E.C.$^{b}$} &
\colhead{$I$([\ion{F}{4}]$_{4060}$)$^c$} &
\colhead{F$^{3+}$/H$^+$} &
\colhead{Ne$^{3+}$/H$^+$}&
\colhead{F/O$^d$} & \colhead{C/O}&
\colhead{O/H} &\colhead{Ref.} \\
\\[-0.09in]
\hline
\\[-0.07in]
IC\,2003 &10    &  1.69$(-3)$& 3.49$(-8)$&  4.19$(-5)$ & 1.65$(-4)$& 0.93& 2.75$(-4)$ & W05b\\
NGC\,2022&12    &  1.84$(-3)$& 2.31$(-8)$&  4.47$(-5)$ & 7.82$(-5)$& 0.46& 4.57$(-4)$ & T03\\
NGC\,2440&10    &  2.17$(-4)$& 3.28$(-9)$&  5.29$(-5)$ & 1.26$(-5)$& 0.90& 4.40$(-4)$ & W05a\\
NGC\,3242&9     &  6.48$(-5)$& 1.75$(-9)$&  9.18$(-6)$:&4.46$(-5)$:& 0.41& 3.31$(-4)$ & T03\\
NGC\,3918&9     &  3.75$(-4)$& 7.84$(-9)$&  5.95$(-5)$ & 1.69$(-5)$& 0.60& 7.24$(-4)$ & T03\\
NGC\,4361&12$^+$&  3.10$(-3)$& 2.57$(-8)$&  2.54$(-5)$ & 6.82$(-4)$& 2.0 & 6.61$(-5)$ & L98\\
NGC\,6302&10    &  6.77$(-4)$& 4.99$(-9)$&  2.37$(-5)$ & 6.35$(-5)$& 0.30& 2.51$(-4)$ & T03\\
NGC\,7027&11    &  9.10$(-4)$& 1.71$(-8)$&  4.19$(-5)$ & 1.04$(-4)$& 2.7 & 4.57$(-4)$ & Z05 \\
NGC\,7662&10    &  8.50$(-4)$& 1.47$(-8)$&  4.30$(-5)$ & 5.93$(-5)$& 1.4 & 3.31$(-4)$ & L04 \\
\enddata     
\tablenotetext{a}{Here $a(-b)=a\times10^{-b}$; 
$^{b}$From \citet{g97}; $^{c}$Corrected for reddening, and on a scale 
where H$\beta=1$; $^{d}$The solar F/O ratio is 5.89$(-5)$ \citep{lodders}.}
\tablerefs{ [B00] \citet{b00}; [L98] \citet{l98}; [L04] \citet{l04}; [P04] \citet{p04}; [S04] \citet{s03}; 
[T03] \citet{t03}; [W05a] \citet{w05a}; [W05b] \citet{w05b}; [Z05] \citet{z05}.
}
\end{deluxetable*}

An unidentified weak feature has been detected in the echelle spectrum of
IC\,418 at 4869\,{\AA} \citep[cf.][]{s03}. We identify it as the [\ion{F}{2}]
$^1$D$_2$--$^3$P$_1$ transition at 4869.02\,{\AA}.  Given the fact that the
[\ion{F}{2}] $\lambda\lambda4789,4869$ lines arise from the same upper level,
the $\lambda4789/\lambda4869$ intensity ratio depends only on their spontaneous
transition probabilities, and not on the nebular physical conditions. For
IC\,418, the measurements yield a $\lambda4789/\lambda4869$ intensity ratio of
5.2.  In the spectrum of the bright high-excitation PN NGC\,7027, we have
detected the [\ion{F}{4}] $^1$D$_2$--$^3$P$_1$ transition at 3996.96\,{\AA}
\citep[cf.][]{z05}. The measurements yield an intensity ratio
$I(4060)/I(3997)=3.6$.  Note that under the assumption of $LS$-coupling, the
intensity ratio of the $^1$D$_2$ -- $^3$P$_2$ to $^1$D$_2$ -- $^3$P$_1$
transitions is simply given by $3\times [\lambda$($^1$D$_2$ --
$^3$P$_1$)/$\lambda$($^1$D$_2$ -- $^3$P$_2$)$]^3$.  This relation predicts an
intensity ratio of 3.2 and 2.8 for [\ion{F}{2}] $\lambda4789/\lambda4869$ and
[\ion{F}{4}] $\lambda4060/\lambda3997$, respectively.  On the other hand, in
the well studied case of the much stronger [\ion{O}{3}]
$\lambda\lambda$4959,5007 nebular lines, based on high quality observations,
\citet{m99} demonstrate that $\lambda5007/\lambda4959 = 3.00\pm0.01$,
significantly higher than the value of 2.88 predicted by the above theoretical
relation under $LS$-coupling.  \citet{s00} have recently recalculated intensity
ratios of the $^1$D$_2$ -- $^3$P$_2$ to $^1$D$_2$ -- $^3$P$_1$ transitions for
carbon-like and oxygen-like ions, taking into account the effects of the
relativistic corrections for the magnetic dipole operators. For [\ion{O}{3}],
they find that the $\lambda5007/\lambda4959$ ratio increases from the
$LS$-coupling value 2.88 to 2.98; the latter is in excellent agreement with the
measurement of \citet{m99}.  For [\ion{F}{2}] and [\ion{F}{4}], the
corresponding line ratios obtained by \citet{s00} are 3.12 and 2.90,
respectively.  Considering the uncertainties in measuring the extremely faint
[\ion{F}{2}] and [\ion{F}{4}] lines, especially the weaker [\ion{F}{2}]
$\lambda$4869 line, which falls very close to the strong H$\beta$ and suffers
from severe contamination by scattered light from H$\beta$, the discrepancies
between the observations and theory, i.e.  5.2 versus 3.12 in the case of
[\ion{F}{2}] in IC\,418, and 3.6 versus 2.90 in the case of [\ion{F}{4}] in
NGC\,7027, are probably insignificant.

Finally, we note that in the infrared, the [\ion{F}{4}] $25.8\mu$m and
$44.2\mu$m fine-structure lines, analogues of the [\ion{O}{3}] $51.8\mu$m and
$88.3\mu$m lines, have intensities comparable to the optical [\ion{F}{4}]
$\lambda\lambda3996,4060$ lines, and are hence good candidates to search for
fluorine in PNe, especially in those that suffer from large dust extinction in the
optical.

\section{Fluorine abundance determinations}

F$^+$/H$^+$ and F$^{3+}$/H$^+$ ionic abundances have been derived from the
observed intensities of the [\ion{F}{2}] and [\ion{F}{4}] lines, respectively,
following standard procedures. Multilevel collisional-radiative models were
constructed for F$^+$ and F$^{3+}$, using collision strengths from \citet{b94}
and \citet{l94}, and radiative transition probabilities from \citet{b88} and
\citet{f85}, respectively. When calculating F$^+$ ionic abundances, we have
adopted electron densities derived from the [\ion{S}{2}]
$\lambda6731/\lambda6716$ ratio and temperatures deduced from the [\ion{N}{2}]
nebular-to-auroral line ratio. For F$^{3+}$, densities derived from the
[\ion{Ar}{4}] $\lambda4740/\lambda4711$ ratio and temperatures obtained from
the [\ion{O}{3}] nebular-to-auroral line ratio were used.  Given the relatively
high critical densities ($>10^6$\,cm$^{-3}$) of the [\ion{F}{2}] $\lambda$4789
and [\ion{F}{4}] $\lambda$4060 lines, the abundances are essentially
independent of the adopted electron densities. For NGC\,4361, we have adopted
the F$^{3+}$/H$^+$ ratio of \citet{l98}, obtained using the same atomic data
set.  The results are presented in Table~\ref{F_PN}.

To convert ionic abundance ratios to those of total elemental abundances,
ionization corrections need to be made for unseen ions.  Given that O$^+$ and
Ne$^{3+}$ have ionization potentials comparable to F$^{+}$ and F$^{3+}$,
respectively, we assume that ${\rm F}/{\rm O}={\rm F}^+/{\rm O}^+$ and ${\rm
F}/{\rm O}=({\rm F}^{3+}/{\rm Ne}^{3+})({\rm Ne}/{\rm O})$, for low- and
high-excitation PNe, respectively. Ionic and elemental abundances of Ne, O and
C adopted in our analysis were taken from the literature.  Table~\ref{F_PN}
summarizes the abundances.  Typical uncertainties in the F/O ratios are
about 12 per\,cent.

\section{Discussion}

\begin{figure}
\epsfig{file=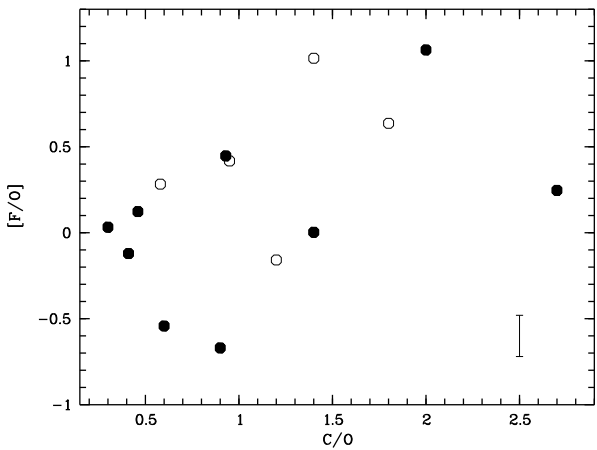,
height=6cm, bbllx=100, bblly=279, bburx=528, bbury=602, clip=, angle=0}
\caption{[F/O] versus C/O. Open and filled circles denote low- and
high-excitation PNe, respectively. A correlation between the two quantities
is apparent. The error bar on the lower right indicates typical uncertainties
of [F/O].
\protect\label{co}}
\end{figure}

\begin{figure}
\epsfig{file=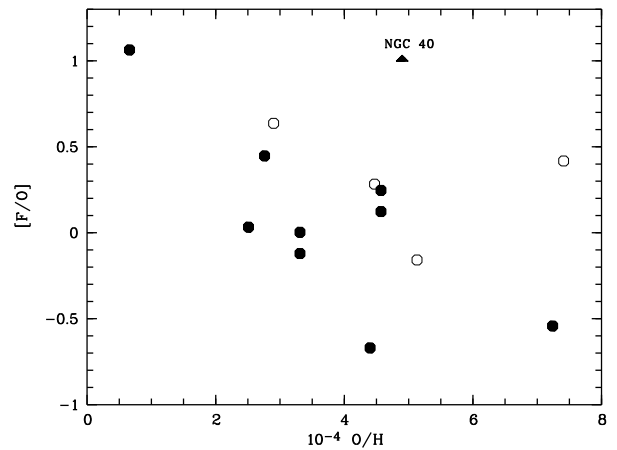, 
height=6cm, bbllx=100, bblly=279, bburx=528, bbury=602, clip=, angle=0}
\caption{[F/O] versus O/H. Open circles and filled circles denote low- and
high-excitation PNe, respectively. With the exception of NGC\,40 (represented
by a filled triangle), an anti-correlation between F/O and O/H is
observed amongst other PNe.
\protect\label{fo}}
\end{figure}

The most recent measurement of the solar F/O abundance ratio yields ${\rm F/O}
= 5.89\times$10$^{-5}$ \citep{lodders}.  Table~\ref{F_PN} shows that most PNe
exhibit F overabundances. The sample yields an average value [F/O]$=$0.3. Our
measurements thus provide further evidence in support of \citet{j92} who
suggest that AGB stars can be a significant source of F production.

According to the AGB evolution models of \citet{j92}, fluorine is synthesized
in the He intershell and then dredged up to the surface during the thermal
pulses. During this process, He is converted into $^{12}$C via partial
He-burning in the intershell and then transported to the stellar surface along
with F.  The scenario thus predicts a positive correlation between the [F/O]
abundance enhancement and the C/O ratio.  Such a correlation is indeed observed
in red giants \citep{j92} and in our current sample of PNe. In Fig.~\ref{co},
we plot observed [F/O] versus C/O ratio for the sample. It is clear that [F/O]
increases with increasing C/O ratio, which strongly suggests that F is indeed
produced in He-burning environments of thermally pulsing AGB stars.
Fig.~\ref{co} also shows that there is no systematic difference between the F
abundances deduced for low- and high-excitation PNe in our sample, suggesting
that the ionization correction formulae adopted in our analysis are probably
reliable.

No PN in our sample shows [F/O] overabundances as much as 30 times solar, as
found by \citet{j92} in giant stars. The highest [F/O] value observed in our
sample is in the halo PN NGC\,4361, which yields a F/O ratio 12 times solar.
\citet{lugaro04} present new calculations of F yields in AGB stars. Their
models fail to match the very high F abundances found by \citet{j92}. A
comparison with Fig.~3 of \citet{lugaro04} shows that the [F/O] and C/O
abundance ratios derived here for the current sample of PNe are in good
agreement with the predictions of their 3\,$M_\sun$ and $Z=0.02$ model. The
very high F abundances observed in some red giants thus remain a puzzle.

A large F enhancement of 10 times solar is observed in NGC\,40, a PN ionized by
a WC8 central star \citep{smith1969,crowther}.  The large enhancement is
probably a consequence of the strong stellar winds from its progenitor star,
leading to a larger amount of the synthesized F being ejected into space. It is
possible that WR PNe play an important role in F production. Further
spectroscopy of additional WR PNe is however required to verify this
possibility.

Fig.~\ref{fo} presents [F/O] versus O/H for the sample.  Excluding the WR PN
NGC\,40, a negative correlation is seen, indicating that the dredge-up process
that produces F is more efficient in a metal-poor environment. The result is in
accord with the predictions of the dredge-up models \citep[e.g.][]{bs88} and
the fact that more carbon stars are observed in metal-poor galaxies
\citep{g04}.  NGC\,40 does not follow this general trend, however
it is consistent with
the idea that strong stellar winds may have dramatically enhanced F in this
nebula (see above).

It has been found that $s$-elements are generally overabundant in PNe
\citep{w05a}, indicating that AGB stars experience strong slow neutron
capture process.  The
origin of the neutrons, however, is much debated \citep[see][for a review]{b99}.
Two possible sources have been proposed, $^{13}$C($\alpha$, n)$^{16}$O and
$^{22}$Ne($\alpha$, n)$^{25}$Mg.  If the later process dominates, then F is
unlikely to survive in the He-intershell since the reaction rate of
$^{19}$F($\alpha$, n)$^{22}$Ne is much higher than that of $^{22}$Ne($\alpha$,
n)$^{25}$Mg \citep[see][for a detailed discussion]{j92}.  Consequently, our
finding that F is generally enhanced in PNe rules out reaction
$^{22}$Ne($\alpha$, n)$^{25}$Mg as the main process of neutron production.

In this paper, we have also presented, for the first time, the F abundance for an
\ion{H}{2} region, the Orion Nebula (see Table~\ref{F_PN}). The value is about
two times higher than those found for K--M dwarfs in the Orion Nebula Cluster
\citep{c05}. The significance of the finding is however questionable, given the
considerable uncertainties in the line measurements -- according to
\citet{b00}, the extremely faint [\ion{F}{2}] $\lambda4789$ nebular line has a
signal-to-noise of only 5.3. Further observation of improved sensitivity should
be helpful.

\section{Conclusions}

We have obtained measurements of the [\ion{F}{2}] $\lambda4789$ and
[\ion{F}{4}] $\lambda4060$ nebular emission lines from the recent literature
for a sample of PNe. The lines are detectable in low- and high-excitation
nebulae, respectively.  From the measurements, abundances of fluorine are
derived.  Our results show that fluorine is overabundant in PNe compared to the
solar value, suggesting that thermally pulsing AGB stars may play an important
role in the production of the element. The result is further corroborated by
the positive correlation observed between the derived [F/O] and C/O. Given that
the WR PN NGC\,40 shows a high F/O ratio, we infer that PNe with a WR central
star are probably significant fluorine producers. Additional observations are
required to verify this point.  

\acknowledgments We thank the anonymous referee for useful comments.  YZ deeply
appreciates R. Williams' hospitality during
the year he worked in STScI. The work is partly supported by NNSFC grant
\#10325312.


\begin{thebibliography}{}

\bibitem[Aller \& Czyzak (1983)]{a83} Aller, L. H., \& Czyzak, S. J.
1983, ApJS, 51, 211

\bibitem[Baldwin et al.(2000)]{b00} Baldwin, J. A. et al.
2000, \apjs, 129, 229 

\bibitem[Baluja \& Zeippen(1988)]{b88} Baluja, K. L., \&
Zeippen, C. J. 1988, J. Phys. B., 21, 1455

\bibitem[Boothroyd \& Sackmann(1988)]{bs88} Boothroyd, A. I., \&
Sackmann, I.-J. 1988, \apj, 328, 671

\bibitem[Busso, Gallino \& Wasserburg (1999)]{b99} Busso, M.,
Gallino, R., \& Wasserburg, G. J. 1999, \araa, 37, 239

\bibitem[Butler \& Zeippen(1994)]{b94} Butler, K., \&
Zeippen, C. J. 1994, \aaps, 108, 1

\bibitem[Crowther et al.(1970)]{crowther} Crowther, P. A., De Marco, O., \& Barlow, M. J. 1998, \mnras, 296, 36


\bibitem[Cunha et al. (2003)]{c03} Cunha, K., Smith, V. V., Lambert,
D. L., \& Hinkle, K. H. 2003, \apj, 126, 1305

\bibitem[Cunha \& Smith (2005)]{c05} Cunha, K., \& Smith, V. V.
2005, \apj, 626, 425

\bibitem[Federman et al. (2005)]{f05} Federman, S. R.,
Sheffer, Y., Lambert, D. L., \& Smith, V. V. 2005, \apj, 619, 884

\bibitem[Fisher \& Saha (1985)]{f85} Fisher, C. F., \& Saha, H. P.
1985, Phys. Scr., 32, 181

\bibitem[Forestini et al. (1992)]{f92} Forestini, M., Goriely, S.,
Jorissen, A., \& Arnould, M. 1992, A\&A, 261, 157


\bibitem[Gurzadyan (1997)]{g97} Gurzadyan, G. A. 1997, The Physics and Dynamics of Planetary Nebulae (Berlin: Springer)


\bibitem[Groenewegen(2004)]{g04} Groenewegen, M. A. T. 2004, in 
Planetary Nebulae beyond the Milky Way, eds.  J. Walsh, L. Stanghellini and N. Douglas (ESO, Garching bei M\"{u}nchen), in press (astro-ph/0407282)



\bibitem[Jorissen et al. (1992)]{j92} Jorissen, A.,  Smith, V. V.,
\& Lambert, D. L. 1992, A\&A, 261, 164


\bibitem[Lennon \& Burke (1994)]{l94} Lennon, D. J., \& Burke, V. M.
1994, \aaps, 103, 273

\bibitem[Liu (1998)]{l98} Liu, X.-W. 1998, \mnras, 295, 699

\bibitem[Liu et al.(2004)]{l04} Liu, Y., Liu, X.-W., Luo, S.-G. \&
      Barlow, M. J. 2004, MNRAS, 353, 1231

\bibitem[Lodders(2003)]{lodders} Lodders, K. 2003, \apj, 591, 1220

\bibitem[Lugaro et al.(2004)]{lugaro04} Lugaro, M. et al.
2004, \apj, 615, 934

\bibitem[Mathis \& Liu (1999)]{m99} Mathis, J. S., \& Liu, X.-W.
1999, \apj, 521, 212

\bibitem[Meynet \& Arnould (2000)]{m00} Meynet, G., \&
Arnould, M. 2000, A\&A, 355, 176


\bibitem[Peimbert et al.(2004)]{p04} Peimbert, M., Peimbert, A.,
Ruiz, M. T., \& Esteban, C. 2004, \apj, 150, 431


\bibitem[Renda et al. (2004)]{r04} Renda, A. et al. 2004, MNRAS, 354, 575


\bibitem[Sharpee et al.(2003)]{s03} Sharpee, B., Williams, R., Baldwin, J. A., \& van Hoof, P. A. M.
2003, \apj, 149, 157

\bibitem[Smith(1969)]{smith1969} Smith, L. F., \& Aller, L. H. 1969, \apj, 157, 1245

\bibitem[Storey \& Zeippen(2000)]{s00} Storey, P. J., \& Zeippen, C. J.
2000, \mnras, 312, 813

\bibitem[Tsamis et al.(2003)]{t03} Tsamis, Y. G., Barlow, M. J., Liu,
X.-W., Danziger, I. J., \& Storey, P. 2003, \mnras, 345, 186


\bibitem[Werner et al. (2005)]{w05} Werner, K., Rauch, T., \& Kruk, J.
W. 2005, A\&A, 433, 641

\bibitem[Wesson et al. (2005)]{w05b} Wesson, R. J., Liu,
X.-W., \& Barlow, M. J. 2005, \mnras, in press

\bibitem[Williams et al. (2005)]{w05a} Williams, R., Zhang, Y., Pellegrini, E., 
Cavagnolo, K., Baldwin, J. A., Sharpee, B., Phyllips, M., \& Liu, X.-W. 2005,
\apj, submitted

\bibitem[Woosley \& Haxton (1988)]{w88} Woosley, S. E., \& Haxton,
W. C. 1988, Nature, 334, 45

\bibitem[Zhang et al(2005)]{z05} Zhang, Y., Liu, X.-W., Luo, S.-G.,
P{\' e}quignot, D., \& Barlow, M. J. 2005, \aap, in press (astro-ph/0507155)

\end{thebibliography}
\end{document}